\documentclass[9pt,shortpaper,twoside,web]{ieeecolor}
\usepackage{generic}
\usepackage{cite}
\usepackage{amsmath,amssymb,amsfonts}
\usepackage{algorithmic}
\usepackage{graphicx}
\usepackage{textcomp}
\usepackage{mathrsfs}
\usepackage{epstopdf}
\usepackage{pifont}

\newcommand{\nn}{\nonumber}
\newcommand{\beq}{\begin{equation}}
\newcommand{\eeq}{\end{equation}}
\newcommand{\bea}{\begin{eqnarray}}
\newcommand{\eea}{\end{eqnarray}}

\def\BibTeX{{\rm B\kern-.05em{\sc i\kern-.025em b}\kern-.08em
    T\kern-.1667em\lower.7ex\hbox{E}\kern-.125emX}}
\begin{document}
\title{Flocking control against the malicious agent}
\author{Chencheng Zhang, 
 Hao Yang, \IEEEmembership{Senior member, IEEE}, Bin Jiang, \IEEEmembership{Fellow, IEEE}, and Ming Cao, \IEEEmembership{Fellow, IEEE}
\thanks{This work was supported in part by the National Natural Science Foundation of China under Grants 62073165 and 62233009, and in part by the Funding of Postgraduate Research $\&$ Practice Innovation Program of Jiangsu Province under Grant KYCX21\underline{~}0222.
\emph{(Corresponding author: Hao Yang)}
}
\thanks{C. Zhang is with the College of Automation Engineering, Nanjing
University of Aeronautics and Astronautics, Nanjing 210016, China,
Institute of Engineering and Technology (ENTEG),  the University of Groningen, the Netherlands (e-mail: zhangchencheng@nuaa.edu.cn).}
\thanks{H. Yang and B. Jiang are with the College of Automation Engineering, Nanjing University of Aeronautics and Astronautics, Nanjing, 211106, China (e-mail: haoyang@nuaa.edu.cn; binjiang@nuaa.edu.cn).}
\thanks{M. Cao is with the Institute of Engineering and Technology (ENTEG),  the University of Groningen, the Netherlands (e-mail: m.cao@rug.nl).
}}

\maketitle

\begin{abstract}
This paper investigates the flocking control of a swarm with a malicious agent that falsifies its controller parameters to cause collision, division, and escape of agents in the swarm. A novel geometric flocking condition is established by designing the configuration of the malicious agent and its neighbors, under which we propose a hierarchical geometric configuration based flocking control method. To help detect the malicious agent, a parameter estimate mechanism is also provided.
The proposed method can achieve the flocking control goal and meanwhile contain the malicious agent in the swarm without removing it. Experimental result shows the effectiveness of the theoretical result.
\end{abstract}


\section{Introduction}
\label{sec:introduction}
\IEEEPARstart{F}{locking} is a form of collective behavior of plenty of interacting agents with a common group objective under limited environmental information and simple rules. Since Reynolds proposed three heuristic rules: separation, alignment, and cohesion for flocking model in \cite{r87}, more and more researchers have put effort into the flocking control problem with its applications in multi-agent systems, mobile agents or networks \cite{r06},\cite{T07}.  The main idea of flocking control is
to make all agents tend to the same velocity and approach a fixed geometric configuration while preserving the swarm connectivity and avoiding collisions
by utilizing artificial intelligence techniques or potential function approaches with local information exchange.
In \cite{Ga11}, a collection of potential functions are designed for swarms of either single or double integrator agents.
Most of these functions are unbounded and are often not appropriate for practice.
Therefore, bounded potential functions are investigated by researchers \cite{wengh12}, \cite{huangjie15}.
What's more, many studies appear in the investigation of swarm intelligence for different tasks. For example, Ref. \cite{G20} considers the aggregation and formation problem with a discrete-time model. In \cite{Bono21}, leader-follower configurations are jointly studied under the model predictive control structure in uncertain environments.

Most of existing results aim at swarm flocking with all agents being healthy and rational. However, agents may suffer from the safety and security issues inevitably in practice.
The misbehavior of a swarm appear largely due to three reasons: faults in the physical
layer \cite{yhTii}, attacks in the cyber layer \cite{D18}, and abnormal/malicious decisions in the supervisory layer \cite{S98}, \cite{D16}.

Under physical faults or cyber attacks, agents may under the appropriate decisions from the supervisory layer. However, malicious decisions refer to the agent's subjectively abnormal and malicious behavior, which are consequences of either  malicious intention or limited cognitive capability of agents.
So the malicious agent in the supervisory layer is more difficult to handle.
Moreover, since the results on flocking under physical faults are already relatively well developed \cite{ya}, \cite{feng}, this paper is devoted to solving flocking problem under abnormal/malicious decisions in the supervisory layer. This is a promising technique with many applications.
A typical example is manned-unmanned multiple (air) vehicle swarm where some malicious members may gain control of vehicles to sabotage the mission of the whole swarm \cite{H18}. Another example is the well-known Byzantine agents who do not obey the prescribed strategy and update their states arbitrarily to threaten the swarm objective \cite{A21}. In real word applications, the control of an Unmanned Aerial Vehicle (UAV) can be taken over by unintended users in a few seconds.

Some effort has been made on control against the malicious agent: For the malicious agent in the cyber layer, the resilient flocking and consensus problems are investigated in \cite{Sa17}-\cite{Ma21}.
In these works, although the malicious agent can communicate untruthful information, they still execute the agreed upon decisions. This makes them quite different from the agent with malicious decisions.
And these researches consider that the malicious agent can be removed
and assume that the network topology remain connected; For the malicious agent in the supervisory layer, Ref. \cite{shang} proposes hybrid $R$-censoring strategies to withstand Byzantine agents and enable cooperative agents to reach consensus.
This approach as well as most of other related results merely relies on excluding the malicious agent.

However, to guarantee the completeness of the task in a swarm, the malicious agent is supposed to be safely contained.
What's more, the above excluding approaches without considering the motions are not applicable for the networked agents subject to geometric or dynamical constraints such as UAV swarms.
To the best of our knowledge, until now almost no result has been reported on \emph{flocking control against malicious decisions of some agent}, let alone the flocking control method that deals with such an agent without excluding it.

Motivated by the above analysis, this paper focuses on the flocking control problem of a swarm in which some agent makes abnormal/malicious decisions in the supervisory layer. Specifically, the malicious agent falsifies its controller parameters, breaks the balance of the attraction or repulsion forces between agents, and thus may lead to collision, division, and escape of agents in the swarm.
As a proverb says ``one rotten apple could ruin a whole barrel of apples", this paper aims at studying \emph{how the malicious agent affects the whole swarm} and \emph{how to achieve the flocking control goal without removing the malicious agent from the swarm}.
The main contributions of this work are summarized as follows:

A novel geometric flocking condition is established to contain the malicious agent by designing the configuration of the malicious agent and its neighbors, under which the forces acting on the malicious agent from its neighbors reach a balance.
We establish a parameter estimate mechanism using filters to help detect the malicious agent with unknown control parameters. Relying on the geometric condition and estimate mechanism, a hierarchical flocking control method is proposed.
Such a method consists of the geometrical configuration based control for the neighbors of the malicious agent and the adaptive flocking control for other normal agents. To the best of our knowledge, this is the first attempt to enable a swarm to against the agent with malicious decisions and achieve the flocking control goal.

The remainder of the paper is organized as follows: In Section II,
preliminaries and model description are given.
Sections III provides the malicious agent containment analysis and the flocking control method.
The experimental result is presented in Section IV, followed by a conclusion in Section V.

\section{Preliminaries}

Notations: Let $\textbf{1}_n$ denote the $n \times 1$ column vector of all ones. Let $|a|_1$ denote the $1$-norm  and $|a|$ denote the Euclidean-norm of $a$, respectively. Let $\textrm{sgn}(a)$ be the signum function of $a$. Let $\textrm{diag}(a_1,\cdot\cdot\cdot,a_p)$ be the diagonal matrix
with diagonal entries $a_1$ to $a_p$. Let $\lambda_{\min}(\cdot)$ denote the minimum eigenvalue of a square real matrix with real eigenvalues. Let $\otimes$ be the Kronecker matrix product.

\subsection{Flocking of a swarm}
Consider a swarm of $N$ agents, whose dynamics take the form
\begin{eqnarray}\label{a1}
\left\{
             \begin{array}{cc}
\dot{x}_i=v_i,&   \\
\dot{v}_i=u_i,& \ \ i\in \mathcal{V},   \end{array}
\right.
\end{eqnarray}
where $x_i\in\Re^{m}$, $v_i\in \Re^{m}$ and $u_i\in \Re^{m}$  denote the position, the velocity  and the control (acceleration) input of agent $i$ for $i\in \mathcal{V}$ with $\mathcal{V}\triangleq\{1,...,N\}$. Define $x_{ij}\triangleq x_i-x_j$ as the relative position between agents $i$ and $j$ for $i,j\in\mathcal{V}$. The model (\ref{a1}) can be transformed from a nonlinear flight control system model \cite{Me99}.

The communication topology between agents in swarm (\ref{a1}) is modeled by an undirected graph $\mathcal{G}=(\mathcal{V},\mathcal{E})$  that consists  of a set of vertices $\mathcal{V}$ and a set of edges $\mathcal{E}\triangleq\{(i,j)|i,j\in \mathcal{V},i\neq j\}$. Vertex $i\in \mathcal{V}$ represents agent $i$, and edge $(i,j)\in\mathcal{E}$ implies that agents $i$ and $j$ can interact with each other and are unordered.
An \emph{undirected path} between vertices $i$  and $j$ is a sequence of unordered edges, $(i , k_1 ), (k_1 , k_2 ),\cdots , (k_l,j)$ with distinct vertices $k_p$, $p=1,2,\cdots,l$.
If there exists an undirected path between vertices $i$ and $j$, the two vertices are said to be \emph{connected}; otherwise, they are \emph{unconnected}.
An undirected graph is called \emph{connected} if any two distinct vertices in the graph are connected.
The Laplacian matrix of graph $\mathcal{G}$ is denoted by $L$.
Define $R$ as the \emph{sensing radius} of each agent, which indicates that two agents can interact only if distance between them is smaller than $R$, i.e., if $0<|x_{ij}|<R$, then $(i,j)\in \mathcal{E}$; otherwise, $(i,j)\notin \mathcal{E}$. Agent $j$ is called a \emph{neighbor }of agent $i$ if $(i,j)\in \mathcal{E}$.
Define $\mathcal{N}(i)\triangleq\{j\in\mathcal{V}: (i,j)\in\mathcal{E},i\neq j\}$  as the set of  neighbors of agent $i$ in $\mathcal{G}$. Note that the following study can be applied to the case that the communication topology is considered static as well.

The flocking control objective is to \emph{make the whole swarm tend to a common speed and approach a fixed configuration without collision, i.e., $\lim_{t\rightarrow\infty}\dot x_{ij}=\lim_{t\rightarrow\infty}v_{i}-v_{j}=0$, $\forall i,j\in \mathcal{V}$; $0\!<\!|x_{ik}(t)|\!<\!R$, $t\geq 0$, $\forall i\in \mathcal{V}$, $k\in \!\mathcal{N}(i)$}.
A conventional flocking control law is designed as \cite{r06}
\bea
u_i=-\sum_{j\in \mathcal{N}(i)}(v_i-v_j)-\sum_{j\in \mathcal{N}(i)}\nabla_{x_i}V_{ij}(|x_{ij}|), \ \ i\in\mathcal{V} \label{a2}
\eea
where the first term corresponds to the desired velocity alignment, and the second term is the gradient of a potential function $V_{ij}$. Note that many existing potential functions with different forms can be applied here in the normal case, for example, the bounded potential function proposed in  \cite{huangjie15}
\begin{align}
V_{ij}(|x_{ij}|)\triangleq\underbrace{\frac{R^2-|x_{ij}|^2}{|x_{ij}|^2+\frac{R^2}{E}}}_{V_{rij}}
+\underbrace{\frac{|x_{ij}|^2}{R^2-|x_{ij}|^2+\frac{R^2}{E}}}_{V_{aij}}, 0\leq |x_{ij}|\leq R \label{E1}
\end{align}
where $E$ is a positive constant. $V_{ij}$ satisfies the following properties

\setlength{\itemsep}{-2pt}
\begin{itemize}
\item   $V_{ij}(|x_{ij}|)=E$ when $|x_{ij}|=0$ or $|x_{ij}|=R$;
    \setlength{\itemsep}{4pt}
\item  $\frac{\partial V_{ij}(|x_{ij}|)}{\partial (|x_{ij}|)}< 0$ when $|x_{ij}|\in(0,\delta)$ and $\frac{\partial V_{ij}(|x_{ij}|)}{\partial (|x_{ij}|)}> 0$ when $|x_{ij}|\in(\delta,R)$, where $\delta\triangleq\frac{\sqrt{2}R}{2}$.
    \setlength{\itemsep}{4pt}
\end{itemize}
\setlength{\itemsep}{2pt}

Physically, the potential can be divided into $V_{ij}\triangleq V_{aij}+V_{rij}$ where $V_{aij}$ and $V_{rij}$ can be viewed as potentials of attraction and repulsion of agent $i$ with respect to agent $j$, respectively.
Obviously, $V_{ij}$ reaches its minimum when $|x_{ij}|=\delta$.
In the unique distance $\delta$, it holds that $\nabla_{x_i}V_{aij}(\delta)+\nabla_{x_i}V_{rij}(\delta)=0$.
In normal case, one can choose $E>\bar Q\triangleq\frac{1}{2}\sum_{i\in \mathcal{V}}v^T_{i}(0)v_{i}(0)+\frac{N(N-1)}{2}\max_{i,j\in \mathcal{V}}\{\bar V_{ij}(|x_{ij}(0)|)\}$ where $\bar V_{ij}(|x_{ij}|)\triangleq\frac{R^2-|x_{ij}|^2}{|x_{ij}|^2}
+\frac{|x_{ij}|^2}{R^2-|x_{ij}|^2}$. This makes the potential between any two agents sufficiently large when the distance between them is equal to 0 or $R$, and thus avoids the collision while preserving the connectivity \cite{huangjie15}.
In the sequel, $E$ will be chosen sufficiently large (i.e., larger than the initial energy functions built in the following sections) to avoid the collision and preserve the connectivity when applying the potential function $V_{ij}$ in the control design.

\subsection{Malicious agent}
Consider a malicious agent, denoted as $i_f\in \mathcal{V}$, who intentionally falsifies controller parameters such that
\begin{eqnarray}
u_{i_f}=-k_v\sum_{j\in \mathcal{N}(i_f)}(v_{i_f}-v_j)-\sum_{j\in \mathcal{N}(i_f)}\nabla_{x_{i_f}}\tilde{V}_{{i_f}j}(|x_{{i_f}j}|)\label{a3}
\end{eqnarray}
where
\begin{eqnarray}
\tilde{V}_{{i_f}j}(|x_{{i_f}j}|)\triangleq k_a V_{a{i_f}j}(|x_{{i_f}j}|)+k_r V_{r{i_f}j}(|x_{{i_f}j}|) \label{aa3}
\end{eqnarray}

We provide some insights on these parameters:
\begin{itemize}
\item   $k_a$: This parameter represents the strength of the attractive force on agent $i_f$ which is inverted for $k_a<0$,  completely lost for $k_a=0$,  partially lost for $0<k_a<1$, and strengthened for $k_a>1$.
    \setlength{\itemsep}{4pt}
\item  $k_r$:  This parameter represents the strength of the repulsive force on agent $i_f$ which is inverted for $k_r<0$, completely lost for $k_r=0$,  partially lost for $0<k_r<1$, and strengthened for $k_r>1$.
    \setlength{\itemsep}{4pt}
\item  $k_v< 1$: This parameter represents the efficacy for the velocity consensus of agent $i_f$ which is inverted for $k_v<0$, completely lost for $k_v=0$, and partially lost for $0<k_v<1$.
    \setlength{\itemsep}{4pt}
\end{itemize}

Compared with the normal controller in (\ref{a2}), the attraction/repulsion effort acting on agent $i_f$ from each of its neighbors is out of balance under the distance $\delta$.
The resultant force of agent $i_f$ is decided by the combination of these three parameters. When $k_v=k_a=k_r=1$, the malicious agent is degenerated into a normal one.

Specifically, there are two circumstances that can cause serious influence to the swarm:  (1) When $k_v\leq0$, $k_a\leq0$ and $k_r\gg1$, the malicious agent $i_f$ tries its best to \emph{run away} from the agents around it and may finally escape from the swarm;
(2) When $k_v\leq0$, $k_r\leq0$ and $k_a\gg1$, agent $i_f$ tries its best to \emph{collide} with the agents around it.

\noindent\textbf{Assumption 1 :} $|k_v|\leq\bar k_v$, $|k_a|\leq\bar k_a$, $|k_r|\leq\bar k_r$ for $\bar k_v, \bar k_a, \bar k_r>0$.
\hspace*{\fill}$\Box$

This assumption means that these parameters are bounded and this is helpful to design the bounds of potential functions.
Such an assumption is not required if the unbounded potential functions instead of the bounded ones are applied in this research.

In the following, the definition of containing a malicious agent is presented.

\noindent\textbf{Definition 1 : }The malicious agent $i_f$ is said to be \emph{contained} if $\dot v_{i_f}=u_{i_f}=0$ and $|x_{i_fj}|=\bar \delta_{i_fj}$  where  $0<\bar\delta_{i_fj}<R$ is a designable expected distance between agent $i_f$ and its neighbor $j\in\mathcal{N}(i_f)$.
\hspace*{\fill}$\Box$

\subsection{Problem formulation}
Define a undirected graph $\mathcal{G}'\triangleq(\mathcal V',\mathcal{E}')$ consisting of the set of vertices $\mathcal V'\triangleq\mathcal{V}-\{i_f\}$ and the set of edges $\mathcal{E}'\triangleq\{(i,j)|i,j\in \mathcal{V}',|x_{ij}|<R, i\neq j\}$.
Define $V_{i}\triangleq\sum_{j\in\mathcal{N}(i)}V_{{i}j}$ for $i\in\mathcal{V},j\in\mathcal{N}(i)$ and $\tilde V_{i_f}\triangleq\sum_{j\in\mathcal{N}(i_f)}\tilde V_{{i_f}j}$.

\noindent\textbf{Assumption 2 :} The initial graph $\mathcal{G}'$ is connected.
\hspace*{\fill}$\Box$

Assumption 2 guarantees that the information among all the normal agents in the swarm can be transmitted  at the initial time.
Similar classical assumption on initial graph can be found in many flocking control researches such as \cite{T07,wengh12}.
Based on this assumption, the problem to be solved is formulated as follows.

\noindent\textbf{Problem $\mathcal{F}$ :} Consider the swarm (\ref{a1}) satisfying Assumptions 1-2 with a malicious agent $i_f\in \mathcal{V}$ under controller (\ref{a3})-(\ref{aa3}).
Design $u_i$, $i\in\mathcal{V}'$ such that

\noindent\ding{172}
$\lim_{t\rightarrow\infty}(v_{i}-v_{j})=0$, \ $\forall i,j\in\mathcal{V}$, i.e., all the agents tend to a same velocity;

\noindent\ding{173} The swarm (\ref{a1}) asymptotically converges to a fixed  geometric  configuration, under which

\begin{itemize}
\item 
$u_{i_f}=0$
and $|x_{i_fj}|=\bar\delta_{i_fj}$,
$\forall j\in\mathcal{N}(i_f)$ where $0<\bar\delta_{i_fj}<R$, i.e., the malicious agent $i_f$ is contained.

\item 
$|x_{ij}|=\tilde{\delta}_{ij}$,  $\forall i\in\mathcal{V}', j\in\mathcal{N}(i)$ where $0<\tilde{\delta}_{ij}<R$, i.e., the normal agents are connected with their neighbors;
\end{itemize}

\noindent\ding{174} $|x_{ij}(t)|\neq 0$  for $t\geq0$, $\forall i,j\in\mathcal{V}$ and $i\neq j$, i.e., no collision occurs.
\hspace*{\fill}$\Box$

\section{Main result}

\subsection{Malicious agent containment analysis}
We first establish a distributed geometric condition under which the malicious agent can still be contained in the swarm. Such a condition will be the basis for the flocking control design.

\noindent\textbf{Lemma 1 : }Consider the swarm (\ref{a1}) with malicious agent $i_f\in \mathcal V$  under controller (\ref{a3})-(\ref{aa3}).
Suppose that $v_{i_f}-v_{j}=0$, $\forall j \in \mathcal{N}(i_f)$.
If
\begin{align}
 |x_{i_fj}|&=\bar\delta, \ \forall j\in\mathcal{N}(i_f),\label{bb1}\\
\sum_{j\in \mathcal N(i_f)}x_{i_fj}&=0  \label{b1}
\end{align}

\noindent\textbf{Proof : }
As $\tilde V_{i_fj}$ defined in (\ref{aa3}) is symmetric with respect to $x_{i_fj}$, it holds that
\begin{align}
\sum_{j\in \mathcal{N}(i_f)}\!\!\nabla_{x_{i_f}}\tilde{V}_{{i_f}j}(|x_{{i_f}j}|)&= \!\!\sum_{j\in \mathcal{N}(i_f)}\!\!\!\nabla_{x_{i_fj}}\tilde{V}_{{i_f}j}(|x_{{i_f}j}|)\nn\\
&=\!\!\sum_{j\in \mathcal{N}(i_f)} \frac{\partial \tilde{V}_{{i_f}j}(|x_{{i_f}j}|)}{\partial |x_{{i_f}j}|} \cdot\frac{\partial |x_{{i_f}j}|}{\partial x_{{i_f}j}} \nn
\end{align}
It yields from the definition of Euclidean norm that for $j\in \mathcal{N}(i_f)$
\begin{align}
\frac{\partial |x_{{i_f}j}|}{\partial x_{{i_f}j}}=\frac{\partial (x_{{i_f}j}^Tx_{{i_f}j})^{\frac{1}{2}}}{\partial x_{{i_f}j}}&=\frac{1}{2}{(x_{{i_f}j}^Tx_{{i_f}j})}^{-\frac{1}{2}}\cdot\frac{\partial (x_{{i_f}j}^Tx_{{i_f}j})}{\partial x_{{i_f}j}}\nn\\
&=\frac{2x_{i_fj}}{{2}{(x_{{i_f}j}^Tx_{{i_f}j})}^{\frac{1}{2}}}=\frac{x_{{i_f}j}}{|x_{{i_f}j}|} \nn
\end{align}
Thus, one can obtain that
\begin{align}
\sum_{j\in \mathcal{N}(i_f)}\!\!\nabla_{x_{i_f}}\tilde{V}_{{i_f}j}(|x_{{i_f}j}|)=\!\!\sum_{j\in \mathcal{N}(i_f)} \frac{\partial \tilde{V}_{{i_f}j}(|x_{{i_f}j}|)}{\partial |x_{{i_f}j}|} \cdot\frac{ x_{{i_f}j}}{|x_{{i_f}j}|} \label{b2}
\end{align}
Define $s$ as the number of agents in $\mathcal{N}(i_f)$. Condition (\ref{bb1}) indicates that $|x_{i_f{j_1}}|=|x_{i_f{j_2}}|=\cdot\cdot\cdot=|x_{i_f{j_s}}|$ for $j_1,j_2,...,j_s \in \mathcal{N}(i_f)$. Therefore, it holds
$\tilde{V}_{{i_f}{j_1}}(|x_{{i_f}{j_1}}|)=\tilde{V}_{{i_f}{j_2}}(|x_{{i_f}{j_2}}|)
=\cdot\cdot\cdot=\tilde{V}_{{i_f}{j_s}}(|x_{{i_f}{j_s}}|)$. It yields from (\ref{b2}) that
\begin{align}
&\!\!\!\sum_{j\in \mathcal{N}(i_f)}\!\!\nabla_{x_{i_f}}\tilde{V}_{{i_f}j}(|x_{{i_f}j}|)\nn\\
&= {s}\frac{\partial \tilde{V}_{{i_f}{j}}(|x_{{i_f}{j}}|)}{\partial |x_{{i_f}j}|}\bigg|_{ |x_{{i_f}j}|=\bar\delta }\cdot\frac{(x_{{i_f}{j_1}}+x_{{i_f}{j_2}}+\cdot\cdot\cdot+ x_{{i_f}{j_s}})}{\bar\delta}\nn
\end{align}
According to condition (\ref{b1}), one further has
\begin{align}
&\!\!\!\sum_{j\in \mathcal{N}(i_f)}\!\!\nabla_{x_{i_f}}\tilde{V}_{{i_f}j}(|x_{{i_f}j}|)\nn\\
&= {s}\frac{\partial \tilde{V}_{{i_f}{j}}(|x_{{i_f}{j}}|)}{\partial |x_{{i_f}j}|}\bigg|_{ |x_{{i_f}j}|=\bar\delta }\cdot\frac{\sum_{j\in \mathcal N(i_f)}x_{i_fj}}{\bar\delta} =0 \nn
\end{align}
Suppose that $ v_{i_f}-v_j=0$ for $j\!\in\!\mathcal{N}(i_f)$. According to the malicious controller (\ref{a3})-(\ref{aa3}) of $i_f$, under conditions (\ref{bb1})-(\ref{b1}), one can deduce that
\begin{align}
\dot v_{i_f}= u_{i_f}&=-\!\!\!\sum_{j\in \mathcal N(i_f)}k_v(v_{i_f}-v_j)-\!\!\!\sum_{j\in \mathcal{N}(i_f)}\!\!\nabla_{x_{i_f}}\tilde{V}_{{i_f}j}(|x_{{i_f}j}|)\nn\\
&=0 \label{f1}
\end{align}
This completes the proof.
\hspace*{\fill}$\Box$

\noindent\textbf{Remark 1 :}
Conditions (\ref{bb1})-(\ref{b1}) provide a desired geometrical configuration that is a regular polygon with the malicious agent being the center and its neighbors being vertexes.
In this case, the total potential gradient of the malicious agent with respect to its neighbors is restricted to be $0$ and their distances are also fixed. Physically, this means that the forces acted on the malicious agent from all its neighbors reach a balance such that the malicious agent can still be contained in the swarm.
An example of the desired configuration satisfying (\ref{bb1})-(\ref{b1}) is presented in Fig. 1, where the malicious agent is surrounded by three neighbors.
\hspace*{\fill}$\Box$

\noindent\textbf{Remark 2 :} Conditions (\ref{bb1})-(\ref{b1}) require the number of agents in $\mathcal{N}(i_f)$, $s\geq 2$.
This is because when agent $i_f$ has at least two neighbors, there exist expected extreme points of  the total potential  $\tilde V_{i_f}$ such that $\nabla_{x_{i_f}} \tilde V_{i_f}$  can be $0$. Then agent $i_f$'s malicious behavior can be contained.
Provided that $s=1$, $\tilde V_{i_f}$ is only related to $|x_{i_fj}|$ for $j\in\mathcal{N}(i_f)$. According to the malicious controller (\ref{a3})-(\ref{aa3}), $\tilde V_{i_f}$ tries to reach its minimum.
However, as $\tilde V_{i_f}$ reaches its minimum, $|x_{i_fj}|$ reaches an unexpected or even dangerous distance, for example, $|x_{i_fj}|=0$ when $k_r=0$ and $k_a\neq0$ in (\ref{b1}). No expected extreme point can be found since $\tilde V_{i_f}$ monotonically increases with respect to $|x_{i_fj}|$. Once $|x_{i_fj}|\neq0$, it holds $\nabla_{x_{i_f}}\tilde V_{i_f}>0$ which leads to the acceleration of agent $i_f$. Thus, the malicious agent can never be contained.
\hspace*{\fill}$\Box$

\begin{figure}[!h]
\begin{center}
{\includegraphics[width=0.22\textwidth]{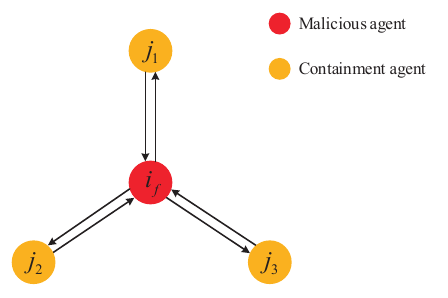}}
\end{center}
\setlength{\abovecaptionskip}{-10pt}
\setlength{\belowcaptionskip}{-10pt}
\caption{An illustration of the desired configuration to contain the malicious agent.}
\vspace{-2mm}
\end{figure}

\subsection {Hierarchical geometric configuration based flocking control}
In this subsection, a hierarchical geometric configuration based flocking control method is proposed to solve problem $\mathcal{F}$.  The control architecture is shown in Fig. 2, where the malicious agent is in Layer 1, all its neighbors are in Layer 2, and other agents in the swarm are in Layer 3.
An important feature of such an architecture is that the agents in Layer 2 do not utilize
the information of agents in Layer 3. This feature makes agents in Layer 2 form the desired configuration as in Lemma 1 more conveniently to contain the malicious one. In this case, the agents in Layers 2 and 3 can be viewed as \emph{leaders} and \emph{followers}, respectively. Define $\mathcal{V}_l$  as the set of agents in Layer 2, $\mathcal{V}_f$ as the set of agents in Layer 3, and $\mathcal{V}_g\triangleq\{i_f\}+\mathcal{N}(i_f)$ as the set of agents in Layers 1 and 2 as shown in Fig. 2. Next we shall design controllers for agents in Layer 2 and Layer 3, respectively.

\begin{figure}[!h]
\begin{center}
{\includegraphics[width=0.3\textwidth]{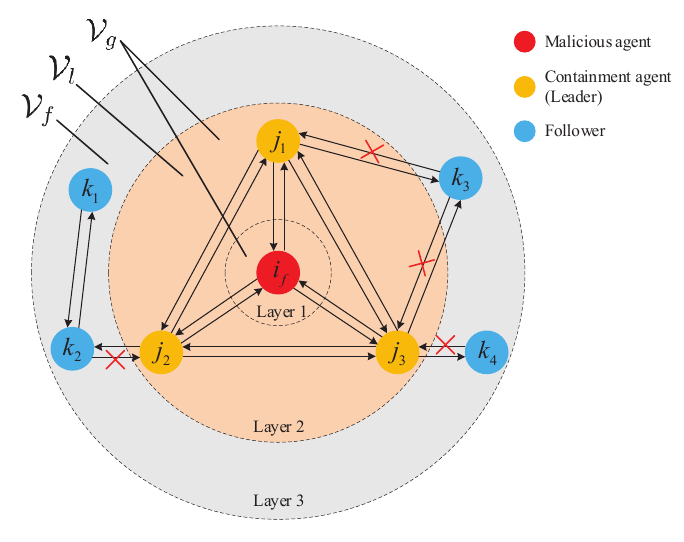}}
\end{center}
\setlength{\abovecaptionskip}{-10pt}
\setlength{\belowcaptionskip}{-10pt}
\caption{An illustration of the hierarchical control architecture.}
\end{figure}

Before giving the main result, the following assumptions are made.

\noindent\textbf{Assumption 3 :} At the initial time, there are at least two neighbors of the malicious agent.
\hspace*{\fill}$\Box$

\noindent\textbf{Assumption 4 :} At the initial time, any two agents in $\mathcal{N}(i_f)$ are neighbors.
\hspace*{\fill}$\Box$

Assumption 3 is a condition on the number of the malicious agent's neighbors
under the geometric configuration method, which has been explained in Remark 2.  Assumption 4 means that all  neighbors of the malicious agent can interact with each other at the initial time.
Such an assumption is needed to resist the influence of the malicious agent by its neighbors jointly under a distributed control structure, and will be explained in details in Remarks 3 and 4.

For convenience, rewrite the dynamics of the malicious agent $i_f$
\begin{align}
\dot v_{i_f}=-C_{i_f} k  \label{dyif}
\end{align}
where $k\triangleq(k_v,k_a,k_r)^T$ is the vector of the unknown parameters and $C_{i_f}\triangleq\left(\sum_{j\in \mathcal{N}(i_f)}(v_{i_f}-v_j),\right.$
$\left.\sum_{j\in \mathcal{N}(i_f)}\nabla_{x_{i_f}}V_{a{i_f}j}(|x_{{i_f}j}|),\sum_{j\in \mathcal{N}(i_f)}\nabla_{x_{i_f}}V_{r{i_f}j}(|x_{{i_f}j}|)\right)$.

In order to track the unknown parameters, filter $v_{i_f}$ and $C_{i_f}$ in (\ref{dyif}) by low-pass first-order filters, one has
\begin{align}
\dot v_{i_f}^F&=-av_{i_f}^F+av_{i_f},\ \  v_{i_f}^F(0)=v_{i_f}(0)\label{filter1}\\
\dot C_{i_f}^F&=-aC_{i_f}^F+C_{i_f}, \ \ C_{i_f}^F(0)=0 \label{filter2}
\end{align}
where $a>0$ is the scalar filter gain. $v_{i_f}^F$ and $C_v^F$ are the filtered $v_{i_f}$ and $C_v$, respectively. They can be obtained by the above stable linear filter equations (\ref{filter1})-(\ref{filter2}).
And it holds that $\dot v_{i_f}^F=-aC_{i_f}^Fk$. This together with (\ref{filter1}) yields
\begin{align}
-v_{i_f}^F+v_{i_f}=-C_{i_f}^Fk \label{filter3}
\end{align}

Define $\hat k\triangleq(\hat k_{v}, \hat k_{a},\hat k_{r})^T$ as the estimate of $k$. Design the adaptive update law of the estimate as follows
\begin{align}
\dot{\hat{k}}=-\Gamma_k C_{i_f}^T\sum_{j\in \mathcal{N}(i_f)}&(v_j-v_{i_f})\nn\\
&-\Gamma_k(C_{i_f}^F)^T(C_{i_f}^F\hat k+v_{i_f}-v_{i_f}^F) \label{estimate}
\end{align}
where
 $\Gamma_k$ is the positive-definite gain matrix.

Based on  Conditions (\ref{bb1})-(\ref{b1}) of Lemma 1, let  $\sum_{j\in \mathcal{N}(i_f)}x_{i_fj}^*=0$ and $|x_{ji_f}^*|=\bar\delta\!<\!R/2$ where $x_{i_fj}^*$ denotes the desired displacement between agents ${i_f}$ and ${j}\in \mathcal{N}(i_f)$.
Note that $x_{jk}\!=x_{ji_f}\!-x_{ki_f}$ and $x_{jk}^*=x_{ji_f}^*\!-x_{ki_f}^*$, $\forall j,k \in\mathcal{N}(i_f)$.  Design the controller of agent $j\in\mathcal{N}(i_f)$ as follows
\begin{align}
u_{j} =& -\kappa_v\!\!\!\!\sum_{p\in \mathcal{N}(j)\cap\mathcal{V}_l}(v_j-v_p)- \kappa_x\!\!\!\!\sum_{p\in \mathcal{N}(j)\cap\mathcal{V}_l}\nabla_{x_j}\hat{V}_{jp} (x_{jp})\nn\\
&-C_{i_f}\hat k \label{b3}
\end{align}
where constants $k_v, k_x\geq1$. The non-negative potential function $\hat{V}_{ij}(x_{ij})$ satisfies the following properties that

\begin{enumerate}
 \item  $\hat V_{ij}$ attains its unique minimum and $\frac{\partial\hat V_{ij}}{\partial |x_{ij}-x^*_{ij}|}=0$ when $x_{ij}=x^*_{ij}$;
     \setlength{\itemsep}{4pt}
 \item  $\hat V_{ij}>\bar H$ when  $|x_{ij}|=0$ and $|x_{ij}|=R$ where $\bar H$ is a designable positive constant.
\end{enumerate}
To solve Problem $\mathcal{F}$, $\bar H$ is chosen as follows
\begin{align}
 \bar H=
     & \!\sum_{j\in\mathcal{N}(i_f)}\bigg(\kappa_x\hat{V}'_{ji_f}(0)
+\frac{\kappa_x}{2}\sum_{i\in\mathcal{N}(i_f)\cap\mathcal{N}(j)}\hat V'_{ji}(0)+\nn\\
     &\frac{1}{2}(v_{j}(0)\!-\!v_{i_f}(0))^T(v_{j}(0)\!-\!v_{i_f}(0))\bigg)\!+\!\frac{1}{2}\lambda_{\max}(\Gamma_k^{-1})\times\nn\\
    &\left((\bar k_v\!+\!\hat k_v(0))^2
+(\bar k_a\!+\!\hat k_a(0))^2\!+ \!(\bar k_r\!+\!\hat k_r(0))^2\right)\label{barH}
\end{align}
where $\hat{V}'_{ij}(0)\triangleq\frac{|x_{ij}(0)-x_{ij}^*|^2}{R-|x_{ij}(0)|}
+\frac{|x_{ij}(0)-x_{ij}^*|^2}{|x_{ij}(0)|}$.

Condition 1) shows that the potential between two agents is minimized when their displacement is equal to the desired one, which makes the two agents approach to the desired configuration. Condition 2) means that the potential would become sufficiently large when the two agents tend to collide or escape, and thus guarantees that  no collision happens and no edge is lost.
One example of such a potential function is as follows
\begin{align}
\hat V_{ij}(x_{ij})\triangleq\frac{|x_{ij}-x_{ij}^*|^2}
{R-|x_{ij}|+\frac{(R-\delta_{ij})^2}{\bar H+\imath}}+\frac{|x_{ij}-x_{ij}^*|^2}
{|x_{ij}|+\frac{\delta_{ij}^2}{\bar H+\imath}}\nn
\end{align}
for $0\!\leq\! |x_{ij}|\!\!\leq \!R$, where $\imath$ is a positive constant and $\delta_{ij}\triangleq|x_{ij}^*|, 0<|x_{ij}^*|<R$.

\noindent\textbf{Remark 3 : }Note that in controller (\ref{b3}), agent $j\in\mathcal{N}(i_f)$ only utilizes the information in $\mathcal{V}_g$ (Layers 1-2) rather than information in $\mathcal{V}_f$ (Layer 3).
In the last term of controller (\ref{b3}), the estimate $\hat k$ of the unknown parameter $k$ with adaptive updating law (\ref{estimate}) is utilized.
And controller (\ref{b3}) requires the state information among all the neighbors of the malicious agent.
This local information exchange is required since all neighbors need to jointly resist the influence of the malicious agent. As will be shown in Theorem 1, under Assumption 4, this local information exchange is always available. We shall explain this setting later in Remark 4.
\hspace*{\fill}$\Box$

Now design a distributed adaptive controller for agent $k\in\mathcal{V}_f$ as
\begin{align}
u_k&=-\!\!\!\!\sum_{p\in \mathcal{N}(k)}\!\alpha_{kp}  \textrm{sgn}(v_k-v_p)-\!\!\!\!\!\sum_{p\in \mathcal{N}(k)}\nabla_{x_k}V_{kp}(|x_{kp}|) \label{u3}\\
\dot \alpha_{kp}&=\gamma_{kp}|v_k-v_p|_{1}, \ p\in \mathcal{N}(k) \nn
\end{align}
where $\alpha_{kp}$ is a varying gain with initial values $\alpha_{kp}(0)\geq 0$ and $V_{kp}$ is defined in (\ref{a2}). $\gamma_{kp}$ is a positive constant and $\gamma_{kp}=\gamma_{pk}$.

\noindent\textbf{Theorem 1 : }Consider the swarm (\ref{a1}) satisfying Assumptions 1-4 with malicious agent $i_f\in \mathcal V$ under controller (\ref{a3})-(\ref{aa3}). Problem $\mathcal{F}$ is solved by applying controller (\ref{b3}) along with parameter estimate update law (\ref{estimate}) to agents in $\mathcal{V}_l$ and controller (\ref{u3}) to agents in $\mathcal{V}_f$. \hspace*{\fill}$\Box$

Before moving on, the following concepts of directed graph theory and a lemma are given that will be used to prove Theorem 1.

A \emph{directed graph} $\hat{\mathcal{G}}$ consists of a pair $(\hat{\mathcal{V}}, \hat{\mathcal{E}})$, where $\hat{\mathcal{V}} \triangleq\{1,\cdots, p\}$ is a set of vertices and $\hat{\mathcal{E}} \triangleq\{(j,k)|j,k\in\hat{\mathcal{V}},j\neq k\} $ is a set of ordered pairs of vertices. An edge $(j,k)$ denotes that vertex $k$ can obtain and utilize information from vertex $j$, but not necessarily vice versa. A \emph{directed path} from vertex $j$ to $k$ is a sequence of edges denoted by $(j, i_1),(i_1 , i_2) ,\cdots,(i_r,k)$ with $i_l\in\hat{\mathcal{V}}, l\in\{1,\cdots,r\}$.

\noindent\textbf{Lemma 2 \cite{Mei} : }For an undirected connected graph with the Laplacian matrix $L\in\Re^{n\times n}$, given $B\triangleq\textrm{diag}(b_1,\cdots,b_n)$ where $b_i\geq 0$, $i = 1,\cdots , n$, if there exists $b_i>0$, then the matrix $E = L + B$ is symmetric positive definite.
\hspace*{\fill}$\Box$

\noindent\textbf{Proof of Theorem 1 : }The proof of Theorem 1 is divided into two parts:
In \emph{Part A}, we prove that $\lim_{t\rightarrow\infty}(v_{i}-v_j)=0$, $\lim_{t\rightarrow\infty}x_{ji}=x_{ji}^*$ and $0<|x_{ij}(t)|<R$ for $t\geq0$, $\forall i,j\in\mathcal{V}_g$.
In \emph{Part B}, we prove that $\lim_{t\rightarrow\infty}(v_{a}-v_b)=0$ and $0<|x_{ab}(t)|<R$ for $t\geq 0$, $\forall a,b\in\mathcal{V}$. $\lim_{t\rightarrow\infty}\nabla_{x_{p}} V_{p}=0$, $\forall p\in \mathcal{V}_f$.

\emph{Part A.} The behavior of agents in Layers 1-2 is considered and the proof is given by analyzing the error velocity and position information between the malicious agent and its neighbors.
Denote $x_f\triangleq(x_{i_f}^T,x_{j_1}^T,...,x_{j_s}^T)^T$, $v_f\triangleq(v_{i_f}^T,v_{j_1}^T,...,v_{j_s}^T)^T$ for $j_{k}\in\mathcal{N}(i_f)$, $k\in\{1,...,s\}$. Define an energy function $H(x_f,v_f)$ as
\begin{align}
H(x_f,v_f)\triangleq &\kappa_x\!\sum_{j\in\mathcal{N}(i_f)}\hat{V}_{ji_f}
+\frac{\kappa_x}{2}\sum_{j\in\mathcal{N}(i_f)}\sum_{i\in\mathcal{N}(i_f)\cap\mathcal{N}(j)}\hat V_{ji}\nn\\
&+\frac{1}{2}\sum_{j\in\mathcal{N}(i_f)}(v_j-v_{i_f})^T(v_j-v_{i_f})\nn\\
&+\frac{1}{2}\tilde k^T\Gamma_k^{-1}\tilde k
\label{H}
\end{align}
where $\tilde k\triangleq(\tilde k_v, \tilde k_a, \tilde k_r)^T$ with $\tilde k_v\triangleq k_v-\hat k_v, \tilde k_a\triangleq k_a-\hat k_a, \tilde k_r\triangleq k_r-\hat k_r$.
Note that $ \dot{\hat V}_{ji}=\dot x_{ji}\nabla _{x_{ji}}\hat V_{ji}$. The time derivative of $H(x_f,v_f)$ is
\begin{align}
\dot H(x_f,v_f)=& \kappa_x\!\!\sum_{j\in\mathcal{N}(i_f)}(v^T_j-v^T_{i_f})\nabla_{x_j}\hat{V}_{ji_f}\nn\\
&
+\frac{\kappa_x}{2}\sum_{j\in\mathcal{N}(i_f)}\sum_{i\in\mathcal{N}(i_f)\cap\mathcal{N}(j)}(v^T_j-v^T_i)
\nabla_{x_j}\hat{V}_{ji}\nn\\
&+\sum_{j\in\mathcal{N}(i_f)}(v_j-v_{i_f})^T(\dot v_j-\dot v_{i_f})
-\tilde k^T\Gamma_k^{-1}\dot{\hat{k}} \nn
\end{align}
Applying the filters (\ref{filter1})-(\ref{filter2}), the estimator (\ref{estimate}) and controller (\ref{b3}), one has
\begin{align}
&\!\!\!\!\dot H(x_f,v_f)\nn\\
=&\kappa_x\!\!\sum_{j\in\mathcal{N}(i_f)}\bigg((v^T_j\!-\!v^T_{i_f})\nabla_ {x_j}\hat{V}_{ji_f}+\frac{1}{2}\!\sum_{i\in\mathcal{N}(i_f)\cap\mathcal{N}(j)}\!(v^T_j-v^T_i)\nn\\
&\times\nabla_{x_j}\hat{V}_{ji}\bigg)
+ \sum_{j\in\mathcal{N}(i_f)}(v_j\!-\!v_{i_f})^T \bigg(-\kappa_v(v_j-v_{i_f})\nn\\
&-\kappa_v\hspace{-4mm}\sum_{p\in \mathcal{N}(i_f)\cap\mathcal{N}(j)}(v_j-v_p)-\kappa_x\hspace{-4mm}\sum_{i\in \mathcal{N}(i_f)\cap\mathcal{N}(j)}\nabla_{x_j}\hat{V}_{ji}\nn\\
&-\kappa_x\nabla_{x_j}\hat{V}_{ji_f} \bigg)-\tilde k^T(C_{i_f}^F)^TC_{i_f}^F\tilde k \label{b5}
\end{align}
It follows from the fact that $x_{ji}=-x_{ij}$ and $x_{ji}-x^*_{ji}=-(x_{ij}-x^*_{ij})$ that
$\frac{\partial\hat V_{ji}}{\partial x_{ji}}=\frac{\partial\hat V_{ji}}{\partial x_{j}}=-\frac{\partial\hat V_{ij}}{\partial x_{i}}$.
Thus, it holds that
\begin{align}
&\frac{\kappa_x}{2}\sum_{j\in \mathcal{N}(i_f)}\sum_{i\in\mathcal{N}(i_f)\cap\mathcal{N}(j)}(v_j^T-v_i^T)\nabla_{x_{j}} \hat V_{ji}\nn\\
=&\frac{\kappa_x}{2}\sum_{j\in \mathcal{N}(i_f)}\sum_{i\in\mathcal{N}(i_f)\cap\mathcal{N}(j)}\Big((v^T_{j}-v^T_{i_f})\nabla_{x_j}\hat V_{ji}\nn\\
&\hspace{1mm}+(v^T_i-v^T_{i_f})\nabla_{x_i}\hat V_{ij}\Big)\nn\\
=&\kappa_x\!\!\!\sum_{j\in \mathcal{N}(i_f)}(v^T_{j}-v^T_{i_f})\sum_{i\in\mathcal{N}(i_f)\cap\mathcal{N}(j)}\nabla_{x_j}\hat V_{ji}  \label{b6}
\end{align}
Combining (\ref{b5}) and (\ref{b6}) yields that
\begin{align}
&\!\!\!\!\dot H(x_f,v_f)\nn\\
=&-\kappa_v\!\!\sum_{j\in\mathcal{N}(i_f)}(v_j-v_{i_f})^T(v_j-v_{i_f})-\tilde k^T(C_{i_f}^F)^TC_{i_f}^F\tilde k\nn\\
&-\!\frac{\kappa_v}{2}\!\!\sum_{j\in\mathcal{N}(i_f)}\sum_{p\in \mathcal{N}(i_f)\cap\mathcal{N}(j)}((v_j\!-\!v_{i_f})\!-\!(v_{p}\!-\!v_{i_f}))^T(v_j\!-\!v_p)\nn\\
=&-\kappa_v\!\!\sum_{j\in\mathcal{N}(i_f)}(v_j
\!-\!v_{i_f})^T(v_j\!-\!v_{i_f})-\!\frac{\kappa_v}{2}\!\sum_{j\in\mathcal{N}(i_f)}\sum_{p\in \mathcal{N}(i_f)\cap\mathcal{N}(j)}\nn\\
&(v_j\!-\!v_p)^T(v_j\!-\!v_p)-\tilde k^T(C_{i_f}^F)^TC_{i_f}^F\tilde k\label{b7}
\end{align}
Therefore, $\dot H(x_f,v_f)$ is always nonpositive
and $H(t)\leq H(0)$ for $t\geq 0$. From the definition of $H(t)$ in (\ref{H}), it holds that $H(t)>\hat V_{ji}(t)$, $\forall i, j\in\mathcal{V}_g$. Thus, $\hat V_{ji}(t)< H(0)$ for $t\geq 0$.
According to the property 2) of $\hat V$, $\hat{V}_{ji}(t)>\bar H$, $\forall j\in \mathcal {N}(i)$ when $|x_{ji}|=0$ and $|x_{ji}|=R$. Since the constant $\bar H$ is chosen as in (\ref{barH}), it holds that $\bar H>H(0)$.
Thus, $\hat{V}_{ji}(t)>\bar H>H(0)$ when $|x_{ji}|=0$ and $|x_{ji}|=R$.
This is contradict to $\hat V_{ji}(t)< H(0), \forall t\geq 0$.
Hence, $|x_{ji}|\neq 0$ and $|x_{ji}(t)|\neq R, \forall t\geq 0$. This guarantees that  the collision  is avoided and no edge is lost between any two agents in $\mathcal{V}_g$.

Define the level set $\Omega_f\triangleq\{(x_f^T,v_f^T)^T\in R^{2(s+1)\times m}: H(t)\leq H(0),H(0)>0\}$. By applying LaSalle's invariance principle, $(x_f^T,v_f^T)^T$ starting in $\Omega_f$ asymptotically converges to the largest invariant set inside the region $\mathcal{C}\triangleq\{(x_f^T,v_f^T)^T\in \Omega_f: \dot H(t)=0\}$. According to (\ref{b7}), $\dot H(t)=0$ holds if and only if $v_1=v_2=\cdot\cdot\cdot= v_{i_f}$ and $C_{i_f}^F\tilde k=0$.
This implies that
$\lim_{t\rightarrow\infty}(v_{i}-v_j)=0$, $\forall i,j\in\mathcal{V}_g$.
and $\lim_{t\rightarrow\infty}C_{i_f}^F\tilde k=0$.
Moreover, according to (\ref{filter2}),  $\lim_{t\rightarrow\infty}C_{i_f}^F={C_{i_f}}/{a}$ where $a>0$.
Thus $\lim_{t\rightarrow\infty}C_{i_f}\tilde k=0$.

In the following, we consider the error system $\dot v_{j}-\dot v_{i_f}=u_{j}-u_{i_f}$ for $j \in\mathcal{N}(i_f)$ at the point $v_1=v_2=\cdot\cdot\cdot= v_{i_f}$.  Obviously, it holds that $\dot v_{j}-\dot v_{i}=0$, $\forall i,j\in \mathcal{V}_g$.
Combining malicious controller (\ref{a3}) and controller (\ref{b3}), one has $\dot v_{j}-\dot v_{i_f}=-\kappa_v\sum_{j\in \mathcal{N}(i_f)}(v_{i_f}-v_j)-\sum_{p\in \mathcal{N}(j)\cap\mathcal{V}_g}(v_j-v_p)- \sum_{p\in \mathcal{N}(j)\cap\mathcal{V}_g}\nabla_{x_j}\hat{V}_{jp}+C_{i_f}\tilde k$. Note that $C_{i_f}\tilde k=0$ at the point $v_1=v_2=\cdot\cdot\cdot= v_{i_f}$ $\forall j\in \mathcal{V}_g$.
Thus, $\sum_{p\in \mathcal{N}(j)\cap\mathcal{V}_g}\nabla_{x_j}\hat{V}_{jp}=0$. Define $\psi_{{\mathcal{G}_f}}(x_f)\triangleq(\cdot\cdot\cdot,|x_{ij}-x_{ij}^*|,\cdot\cdot\cdot )^T$ with $i,j\in \mathcal{V}_g$. Consider the error system in the compact form, one obtains that $R_{{\mathcal{G}_f}}^T(x_f)\xi(x_f)=0$ where $\xi(x_f)\triangleq(\cdot\cdot\cdot, {\partial\hat V_{ij}}/{\partial|x_{ij}-x^{*}_{ij}|},\cdot\cdot\cdot)$ and $R_{{\mathcal{G}_f}}^T(x_f)\triangleq\nabla x_f \psi_{{\mathcal{G}_f}}(x_f) $ is the rigidity matrix.
Since $Rank(R_{{\mathcal{G}_f}}(x_f))=md-m(m+1)/2$ where $m$ is the dimension and $d$ is the vertex number of $\mathcal{V}_g$,
it follows from the Rigidity Theory in \cite{sun18} that $R_{{\mathcal{G}_f}}^T(x_f)\xi(x_f)=0$  is equivalent to $\xi(x_f)=0$.
From the property 1) of $\hat V_{ij}$, we can deduce that ${\partial\hat V_{ij}}/{\partial|x_{ij}-x^{*}_{ij}|}=0$ is equivalent to $|x_{ij}-x^{*}_{ij}|=0$, $\forall i,j\in\mathcal{V}_g$.
Hence, it holds that $x_{ij}\rightarrow x^{*}_{ij}$ as $t\rightarrow \infty$.
Also, it yields from controllers (\ref{a3}) and (\ref{b3}) that  $\dot v_{i_f}=u_{i_f}\rightarrow0$ and $\dot v_{j}=u_{j}\rightarrow0$ for $j\in \mathcal{N}(i_f)$.

\emph{Part B.}
As is shown in Fig. 2, all agents in Layer 2 can be viewed as the leaders of agents in Layer 3.
Let $\bar{\mathcal{G}}$ be the direct graph characterizing the information interaction among agents in $\mathcal{V}_f$ and the transmission from agents in $\mathcal{V}_l (\mathcal{N}(i_f))$ to  agents in $\mathcal{V}_f$.
If there exists a directed path from  agent $j\in\mathcal{N}(i_f)$ to agent $k\in\mathcal{V}_f$ in graph $\bar{\mathcal{G}}$, agent $j$ is said to be a leader of agent $k$.
Here, we prove that leader-follower flocking for agents in Layers 2-3 can be realized under controller (\ref{u3}) by analyzing the graph corresponding to each leader and all its followers.
This together with the results in \emph{Part A} yields that all the followers tend to the same velocity as that of the leaders.

Denote $\mathcal{L}(k)$ as the set of agent $k$'s leaders. Define the energy function $\Upsilon({x},{v})$  as
\begin{align}
\Upsilon({x},{v})
\triangleq& H(x_f,v_f)+ \sum_{i\in\mathcal{V}_f}\sum_{j\in \mathcal{L}(i)\cap\mathcal {N}(i)} {V}_{ij}\nn\\
&+\!\frac{1}{2}\sum_{i\in \mathcal{V}_f}\sum_{p\in\mathcal{V}_f\cap\mathcal{N}(i)}s_{\mathcal{L}(i)}V_{ip}\nn\\
&+\frac{1}{2}\sum_{i\in \mathcal{V}_f} \sum_{j\in \mathcal{L}(i)}(v_i- v_{j})^T(v_i- v_{j})
\nn\\
&+\sum_{i\in\mathcal{V}_f}\sum_{j\in\mathcal{N}(i_f)\cap\mathcal{N}(i)}
\frac{1}{2{\gamma_{ij}}}(\alpha_{ij}-\bar{\alpha})^2\nn\\
&+\sum_{i\in\mathcal{V}_f}\sum_{p\in\mathcal{V}_f\cap\mathcal{N}(i)}
\frac{s_{\mathcal{L}(i)}}{4{\gamma_{ip}}}(\alpha_{ip}-\bar{\alpha})^2 \label{b14}
\end{align}
where $s_{\mathcal{L}(i)}$ denotes the number of agents in set $\mathcal{L}(i)$. Constant $\bar{\alpha}$ will be designed later.

Note that the graph of agents in Layer 3 is undirected, thus $s_{\mathcal{L}(i)}=s_{\mathcal{L}(j)}$, $\forall i\in\mathcal{V}_f, j\in\mathcal{N}(i)\cap\mathcal{V}_f$.
Therefore, the derivative of $\Upsilon$ is
\begin{align}
\dot\Upsilon=&
\dot H
+\sum_{i\in\mathcal{V}_f}\sum_{j\in \mathcal{L}(i)\cap\mathcal{N}(i)}\dot x_{ij}^T\nabla_{x_{ij}}{V}_{ij}
+\frac{1}{2}\sum_{i\in \mathcal{V}_f}\sum_{p\in\mathcal{V}_f\cap\mathcal{N}(i)}\nn\\
&s_{\mathcal{L}(i)}\dot { x}_{ip}^T\nabla_{x_{ip}}V_{ip}
+\sum_{i\in \mathcal{V}_f} \sum_{j\in \mathcal{L}(i)}(v_i- v_{j})^T(\dot v_i- \dot v_{j})
\nn\\
&+\sum_{i\in\mathcal{V}_f}\sum_{j\in\mathcal{N}(i_f)\cap\mathcal{N}(i)}
(\alpha_{ij}-\bar{\alpha})| v_{i}-v_{j}|_1\nn\\
&+\sum_{i\in\mathcal{V}_f}\sum_{p\in\mathcal{V}_f\cap\mathcal{N}(i)}
\frac{s_{\mathcal{L}(i)}}{2}(\alpha_{ip}-\bar{\alpha})|v_{i}-v_{p}|_1\nn
\end{align}
Applying controller (\ref{u3}), one obtains that
\begin{align}
\dot\Upsilon
=&\dot H(x_f,v_f)+\sum_{i\in\mathcal{V}_f}\sum_{j\in \mathcal{L}(i)\cap\mathcal{N}(i)}(v_i-v_j)^T\nabla_{x_{ij}}{V}_{ij}
\nn\\
&+\!\frac{1}{2}\sum_{i\in \mathcal{V}_f}\!\sum_{p\in\mathcal{V}_f\cap\mathcal{N}(i)}\!\!\!\!s_{\mathcal{L}(i)}\dot { x}_{ip}^T\nabla_{x_{ip}}V_{ip}
+\!\!\!\sum_{i\in \mathcal{V}_f} \!\sum_{j\in \mathcal{L}(i)}\!(v_i\!-\! v_{j})^T
\nn\\
&\times\!\bigg(-\hspace{-4mm}\sum_{j\in  \mathcal{L}(i)\cap\mathcal{N}{(i)}}\!\!\!\!\alpha_{ij}\textrm{sgn}(v_i\!-\!v_j)-\hspace{-4mm}\sum_{p\in \mathcal{V}_f\cap\mathcal{N}{(i)}}\!\!\!\!\alpha_{ip}\textrm{sgn}(v_i\!-\!v_p)\nn\\
&
-\sum_{j\in \mathcal{L}(i)\cap\mathcal{N}(i)}\nabla_{x_{ij}}{V}_{ij}-\sum_{p\in \mathcal{V}_f\cap\mathcal{N}(i)}\nabla_{x_{ip}}{V}_{ip}-u_j\bigg)
\nn\\
&+\sum_{i\in\mathcal{V}_f}\sum_{j\in\mathcal{N}(i_f)\cap\mathcal{N}(i)}
(\alpha_{ij}-\bar{\alpha})| v_{i}-v_{j}|_1\nn\\
&+\sum_{i\in\mathcal{V}_f}\sum_{p\in\mathcal{V}_f\cap\mathcal{N}(i)}
\frac{s_{\mathcal{L}(i)}}{2}(\alpha_{ip}-\bar{\alpha})|v_{i}-v_{p}|_1\nn
\end{align}
For convenience, label the agents in $\mathcal{N}(i_f)$ who have neighbors in $\mathcal{V}_f$ as $1$ to $\omega$. If there exists a directed path from agent $j$, $j\in\{1,\cdots,\omega\}$ to some agents in $\mathcal{V}_f$, denote the set of these agents as $F(j)$. Note that $F(j)\subseteq\mathcal{V}_f$, and the leaders of $k,p$ are same if $k,p\in\mathcal{V}_f$ are neighbors.
Therefore, it yields that
\begin{align}
\dot\Upsilon =&\dot H
+\frac{1}{2}\sum_{i\in \mathcal{V}_f}\sum_{p\in\mathcal{V}_f\cap\mathcal{N}(i)}s_{\mathcal{L}(i)}\dot { x}_{ip}^T\nabla_{x_{ip}}V_{ip}
\nn\\
&-\hspace{-1mm}\sum_{i\in \mathcal{V}_f} \sum_{j\in \mathcal{N}(i_f)\cap\mathcal{N}{(i)}}\alpha_{ij}|v_i-v_j|_1
-\hspace{-1mm}\sum_{j=1}^{\omega}\sum_{i\in F(j)} \hspace{-1mm}(v_i-v_j)^T
\nn\\
&
\times\sum_{p\in F(j)\cap\mathcal{N}{(i)}}\alpha_{ip}\textrm{sgn}((v_i-v_j)-(v_p-v_j))\nn\\
&-
\sum_{j=1}^{\omega}\sum_{i\in F(j)} (v_i-v_j)^T
\sum_{p\in \mathcal{V}_f\cap\mathcal{N}(i)}\nabla_{x_{ip}}{V}_{ip}
\nn\\
&-\hspace{-1mm}\sum_{i\in\mathcal{V}_f}\sum_{j\in\mathcal{N}(i_f)\cap\mathcal{N}(i)}\hspace{-3mm}(v_i-v_j)^Tu_j
-\hspace{-1mm}\sum_{j=1}^{\omega}\sum_{i\in F(j)} (v_i-v_j)^Tu_j
\nn\\
&+\sum_{i\in\mathcal{V}_f}\sum_{j\in\mathcal{N}(i_f)\cap\mathcal{N}(i)}
(\alpha_{ij}-\bar{\alpha})| v_{i}-v_{j}|_1
\nn\\&+\sum_{j=1}^{\omega}\sum_{i\in F(j)}(v_{i}-v_{j})\sum_{p\in F(j)\cap\mathcal{N}{(i)}}(\alpha_{ip}-\bar \alpha)\nn\\
&\times\textrm{sgn}\left((v_{i}-v_{j})-(v_{p}-v_{j})\right)\label{r1}
\end{align}
Since  $V_{ip}$ is symmetric with respect to $x_{ip}$ and $x_{ip}=(x_{i}-\chi)-(x_{p}-\chi)$, $\forall \chi\in\Re^{m}$, it holds that
\begin{align}
&\frac{1}{2}\sum_{i\in \mathcal{V}_f}\sum_{p\in\mathcal{V}_f\cap\mathcal{N}(i)}\dot { x}_{ip}^T\nabla_{x_{ip}}V_{ip}(x_{ip})\nn\\
=&\frac{1}{2}\sum_{j=1}^{\omega}\sum_{i\in F(j)}\sum_{p\in F(j)\cap\mathcal{N}{(i)}}((v_{i}-v_{j})-(v_{p}-v_{j}))\nn\\
&\times\nabla_{x_{ji}} \hat V_{ip}((x_{i}-x_{j})-(x_{p}-x_{j}))\nn\\
=&\sum_{j=1}^{\omega}\sum_{i\in F(j)}(v_{i}-v_{j})\sum_{p\in F(j)\cap\mathcal{N}{(i)}}\nabla_{x_{ip}}\hat V_{i}(x_{ip}) \label{r2}
\end{align}
Since $u_{j}(t)$ is continuous for $t\in[0,\infty)$ and it is proved  in \emph{Part A} that $\lim _{t\rightarrow\infty}u_{j}=0$, it holds that $u_j(t)$ is bounded for $t\in[0,\infty)$. Denote the bound as $\mu$ such that $|u_j(t)|_1\leq \mu, \forall j\in\mathcal{N}(i_f)$. Substituting (\ref{r2}) into (\ref{r1}) yields
\begin{align}
\dot\Upsilon\leq&\dot H
+\sum_{i\in\mathcal{V}_f}\sum_{j\in\mathcal{N}(i_f)\cap\mathcal{N}(i)}|u_j||v_i-v_j|\nn\\
&+\sum_{j=1}^{\omega}\sum_{i\in F(j)}|u_j||v_i-v_j|
-\sum_{i\in \mathcal{V}_f} \sum_{j\in \mathcal{N}(i_f)\cap\mathcal{N}{(i)}}\!\!\!\!\bar \alpha|v_i-v_j|_1\nn\\
&
-\sum_{j=1}^{\omega}\sum_{i\in F(j)}\hspace{-2mm} (v_i-v_j)^T\!\!\!\!\!\!\!\sum_{p\in F(j)\cap\mathcal{N}{(i)}}\hspace{-4mm}\bar\alpha\textrm{sgn}((v_i-v_j)\!-\!(v_p-v_j))\nn
\end{align}
Define the number of agents in $F(j)$, $j\in\{1,\cdots, \omega\}$ as $s_{F(j)}$.
Define $\check{v}_j$ as a column stack vector of $(v_{i}-v_{j})$, $i\in F(j)$.
Let $\mathcal{G}_j$ be the undirected  graph characterizing the interaction among the $s_{F(j)}$
followers of leader $j$ with the associated Laplacian matrix $L_{j}\triangleq D_{j}D_{j}^T$. Note that by
definition, $L_{j}$ is symmetric positive semi-definite.
Let $\bar{\mathcal{G}}_j$ be the directed graph characterizing the interaction
among leader $j$ and its followers.
Let the edge weight $a_{ij} = 1$ if leader $j$ is a neighbor of follower
$i$ and $a_{ij} = 0$ otherwise. Define $\Lambda_{j}\triangleq\textrm{diag}(a_{i_1j}, \cdots, a_{i_{s_{F(j)}j}})$, $i_k\in F(j)$, $k=1,\cdots,s_{F(j)}$. Note that $\Lambda_{j}^2=\Lambda_{j}$ because $a_{ij},i\in F(j)$ is either $1$ or $0$.  Therefore, it holds
\begin{align}
\dot\Upsilon \leq&\dot H
+\sum_{j=1}^{\omega}\textbf{1}_{s_{F(j)}}\otimes\mu|\check v_j|-\sum_{j=1}^{\omega}\bar{\alpha}|\Lambda_{j}\otimes I_m\check v_j|_1\nn\\
&-\sum_{j=1}^{\omega}\bar{\alpha}|D^T_{j}\otimes I_m\check v_j|_1\nn
\end{align}
Define the leader-follower topology matrix associated with graph $\bar{ \mathcal{G}}_j$ as $R_j\triangleq L_{j}+\Lambda_{j}$, $j\in\{1,\cdots,\omega\}$.  According to Lemma 2, $R_{j}$ is symmetric positive definite. Based on (\ref{b7}) and the fact that $|\cdot|\leq|\cdot|_1$ for any vector, one obtains
\begin{align}
\dot\Upsilon
\leq&-\kappa_v\sum_{j\in\mathcal{N}(i_f)}(v_j-v_{i_f})^T(v_j-v_{i_f})\nn\\
&-\frac{\kappa_v}{2}\sum_{j\in\mathcal{N}(i_f)}\sum_{p\in \mathcal{N}(i_f)\cap\mathcal{N}(j)}(v_j-v_p)^T(v_j-v_p)\nn\\
&-\tilde k^T(C_{i_f}^F)^TC_{i_f}^F\tilde k-\sum_{j=1}^{\omega}(\bar {\alpha }\sqrt{\lambda_{\min}(R_j)}-\bar \mu)|\check v_{j}|\label{lya}
\end{align}
where $\bar \mu\triangleq\max_{k\in\{s_{F(1)},s_{F(2)},\cdots,s_{F(\omega)}\}}\{\textbf{1}_{k}\otimes\mu\}$. If $R_j(t)$, changes at some time, there exists $t_1 > 0$ such that $R_j(t)=R_j(0)$, $\forall t\in[0,t_1)$. By designing $\bar {\alpha }> {\bar \mu}/{\min_{j\in\{1,\cdots,\omega\}}\{\sqrt{\lambda_{\min}(R(j)(0))}\}}$, one has $\dot\Upsilon(t)\leq 0$ for $t\in[0,t_1)$. Since $V_{ik}(t)$, $i\in\mathcal{V}'$, $k\in\mathcal{N}(j)\cap \mathcal{V}'$ is continuous, we can conclude that $V_{ik}(t_1)\leq \Upsilon(t_1)$.
From the definition of $V_{ik}$ in (\ref{a2}), it follows that there is no collision and no edge in the graph $\bar{\mathcal{G}}_j(0)$ will be lost for $t \in[0, t_1]$. Therefore, the only possibility that $R_j(t)$ changes at
$t = t_1$ is that some edges are added in the graph, which means that $\bar{ \mathcal{G}}_j(0)$ is a subgraph of $\bar{ \mathcal{G}}_j(t_1)$. Then it yields from Lemma 2.2 in \cite{Sheida} that
$R_j(0)\leq R_j(t_1), \forall j\in\{1,\cdots,\omega\}$ and thus $\lambda_{\min}(R_j(0)) \leq \lambda_{\min}(R_j(t_1))$.
Following the same argument, if $R_j(t)$ changes at $t = t_i > t_1$, $i=1,\cdots$, one can have the same conclusion. Therefore, it holds that $ {\bar \mu}/{\min_{j\in\{1,\cdots,\omega\}}\{\sqrt{\lambda_{\min}(R(j)(0))}\}}\geq {\bar \mu}/{\min_{j\in\{1,\cdots,\omega\}}\{\sqrt{\lambda_{\min}(R(j)(t))}\}}$ for all $t\geq 0$. And there is no collision and also no edge in the graph $\bar{\mathcal{G}}_j(0)$ is lost for all $t\geq 0$.
Thus, $\dot\Upsilon(t)\leq 0$, $\forall t\geq 0$.

Combining (\ref{lya}) with the analysis in \emph{Part A}, it holds that $\lim_{t\rightarrow\infty}(v_{j}-v_{i_f})=0$ and $\lim_{t\rightarrow\infty}(v_{i}-v_{j})=0$, $\forall i\in F(j),j\in\{1,\cdots,\omega\}$. Assumption 2
indicates that there exists at least one leader in $\{1,\cdots,\omega\}$ for each agent in $\mathcal{V}_f$ when $t=0$. Since no edge in $\bar{ \mathcal{G}}_j$, $\forall j\in\{1,\cdots, \omega\}$ is lost for $t\geq 0$, all agents in $\mathcal{V}_f$ are followers of agents in $\{1,\cdots, \omega\}$  for $t\geq 0$.
This further leads to $\lim_{t\rightarrow\infty}(v_{i}-v_{j})=0, \forall i,j\in\mathcal{V}$. This completes the proof.
\hspace*{\fill}$\Box$

\noindent\textbf{Remark 4 : }According to the proof of Theorem 1, $\lim_{t\rightarrow\infty}(x_{ij}-x^*_{ij})=0$, $\forall i,j\in\mathcal{V}_g$ holds. Since $|x^*_{i_fj}|=\bar\delta_{i_fj}<R/2$, $\forall j\in\mathcal{N}(i_f)$, it holds $|x^*_{pk}|=|x^*_{pi_f}+x^*_{i_fk}|\leq|x^*_{i_fp}|+|x^*_{i_fk}|<R$, $\forall p,k\in\mathcal{N}(i_f)$
Also, no edge in $\mathcal{V}_g$ is lost under the controller (\ref{b3}). This together with Assumption 4 that any two agents in $\mathcal{N}(i_f)$ are neighbors at the initial time guarantees that any two agents in $\mathcal{N}(i_f)$ are always neighbors for all $t\geq 0$.
Therefore, the local information exchange among all the neighbors of the malicious agent can be obtained as they can interact with each other.
If Assumption 4 is not satisfied,  this information  can be achieved in virtue of a local communication network among agents in Layer 2 \cite{yangh15}. Such a local network can be built conveniently if it does not exist, since these agents are close to each other.  With this local network, $\bar\delta_{i_fj}$ can be any desired value in $(0,R)$.
\hspace*{\fill}$\Box$

\noindent\textbf{Remark 5 : }The main idea of the geometric configuration control  (\ref{b3}) is to contain the malicious agent by ``pulling" its neighbors to the desired geometric shape. In the controller, the first term is to urge the agents to reach the same common velocity. The second term is to let the agents approach to the desired configuration to contain the malicious agent. The last term is to compensate for the influence of the malicious agent reacting on its neighbors.
\hspace*{\fill}$\Box$

\section{Experimental result}
In this section, the experimental result is presented to illustrate the effectiveness of the proposed flocking control scheme in the above section.

A semi-physical experimental platform of an Unmanned Aerial Vehicle(UAV) swarm has been set up based on 40 Raspberry Pi computers. The dynamics and controller of each UAV are simulated by 2 Raspberry Pi computers, respectively.
Specifically, the flight control system model of UAV in the platform and the transformation method between the UAV model and model (1) are from Ref. \cite{Me99}.
Fig. 3 is the picture of the UAV swarm semi-physical platform, which consists of 4 parts: Raspberry Pi computers, a thrust lever, a data analysis terminal and a flight display terminal.

In the experiment, we consider a 2-dimensional swarm of 13 UAVs (UAVs 0-12), including a malicious agent (UAV 6) under controller (\ref{a3})-(\ref{aa3}) with $k_a=0, k_r=450000$ and $k_v=0.8$.
Define the velocity of UAV $i\in\{0,1,...,12\}$ as $v_i\triangleq (v_{\textrm{x}i}, v_{\textrm{y}i})$ where $v_{\textrm{x}i}$ and $v_{\textrm{y}i}$ are velocities in \textrm{x}-dimension and \textrm{y}-dimension, respectively.
The control inputs of UAV $i$ is the banking angle $\Phi_i$, lift $L_i$ and engine thrust $T_i$. The initial ground velocity $\textrm{V}_{i},i\in\{0,1,...,12\}$ of UAV $i$ is taken randomly from $(27,35)m/s$ and heading angle $\chi_i$ is taken from $(\pi/6,\pi/4)\textrm{rad}$.  The initial flight path angle is $0$.
According to the model transformation in \cite{Me99}, $v_{\textrm{x}i}\triangleq \textrm{V}_i\cos(\gamma_i)\cos\chi_i$ and $v_{\textrm{y}i}\triangleq \textrm{V}_i\cos(\gamma_i)\sin\chi_i$. Let the communication distance be $R=18\sqrt{2}\textrm{m}$, thus $\delta=18\textrm{m}$. Let the desired distance between the malicious agent and its neighbors be $\bar\delta=12\textrm{m}<R/2$.
Apply  controller (\ref{b3}) with $\kappa_v=6$ and $\kappa_x=2$ to  UAVs 2, 5, 7 and 10. Apply controller (\ref{u3}) with $\gamma_{kp}=1$ to UAVs 0, 1, 3, 4, 8, 9, 11 and 12. The experimental result presented in Figs. 4-5 shows that all UAVs tend to a common velocity and all the control efforts tend to $0$.
The malicious UAV 6 is contained, and the distances between it and its neighbors tend to $12\textrm{m}$ as expected and the configuration tends to the desired one as is shown in Fig. 6.

\begin{figure}[!ht]
\vspace{-0.3cm}
\centering{
\includegraphics[width=0.4\textwidth]{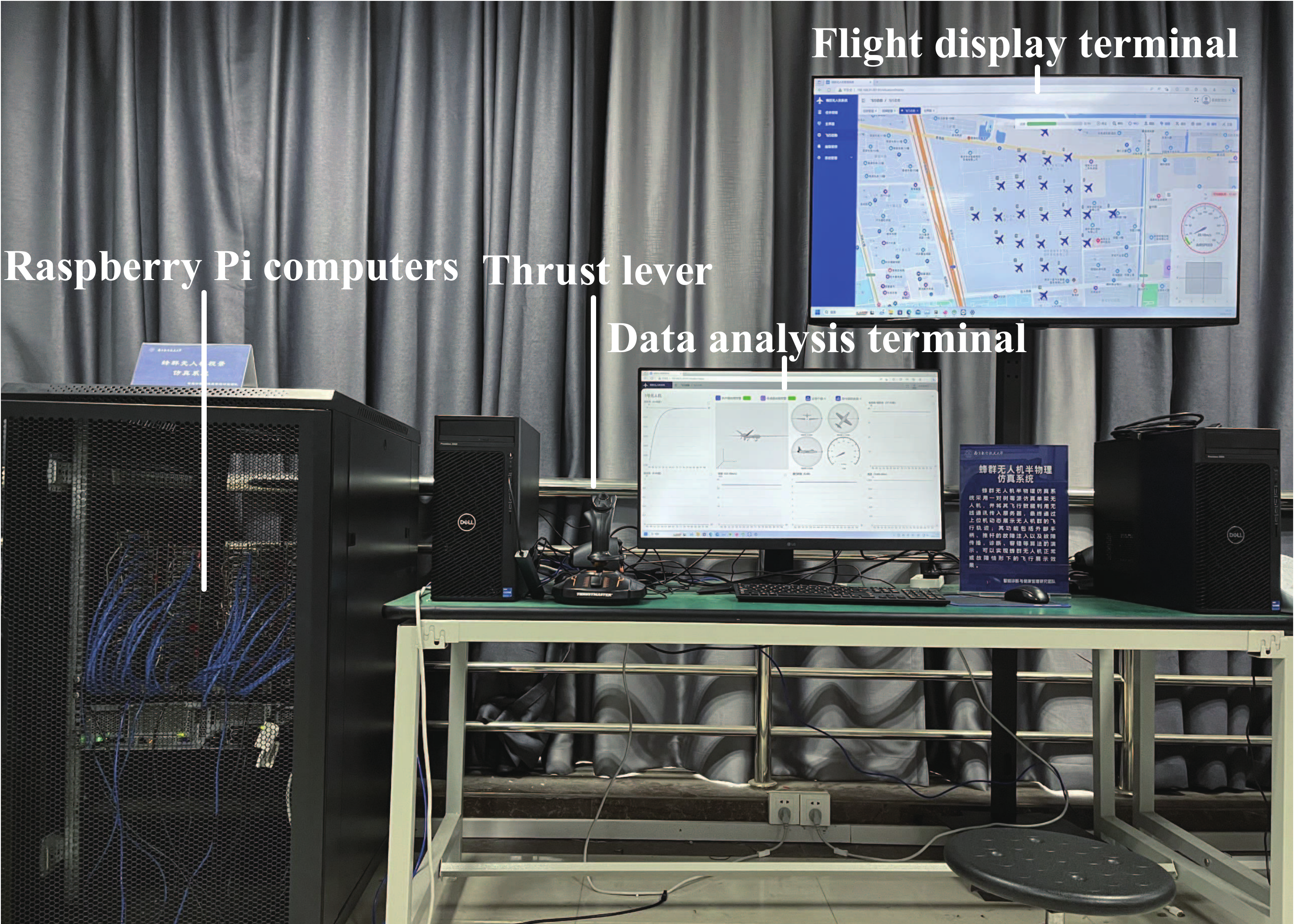}}
\caption{Overview of UAV swarm semi-physical experimental platform.}
\end{figure}

\begin{figure}[!ht]
\vspace{-0.3cm}
\centering{
\hspace{-1cm}\includegraphics[width=0.71\textwidth]{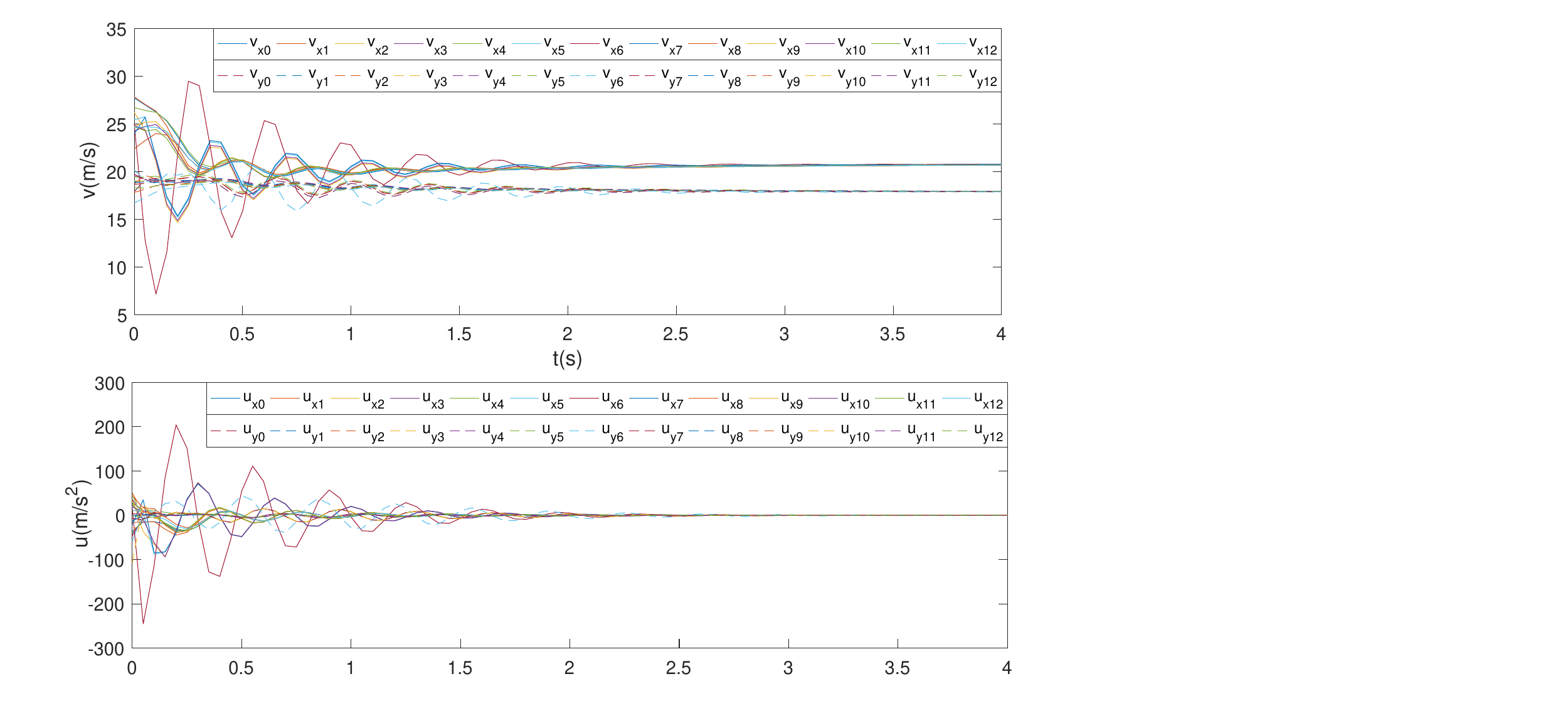}}
\setlength{\abovecaptionskip}{-8pt}
\setlength{\belowcaptionskip}{-8pt}
\caption{Trajectories  of velocities and control efforts of UAVs 0-12.}
\end{figure}

\begin{figure}[!ht]
\vspace{-0.3cm}
\centering{
\includegraphics[width=0.5\textwidth]{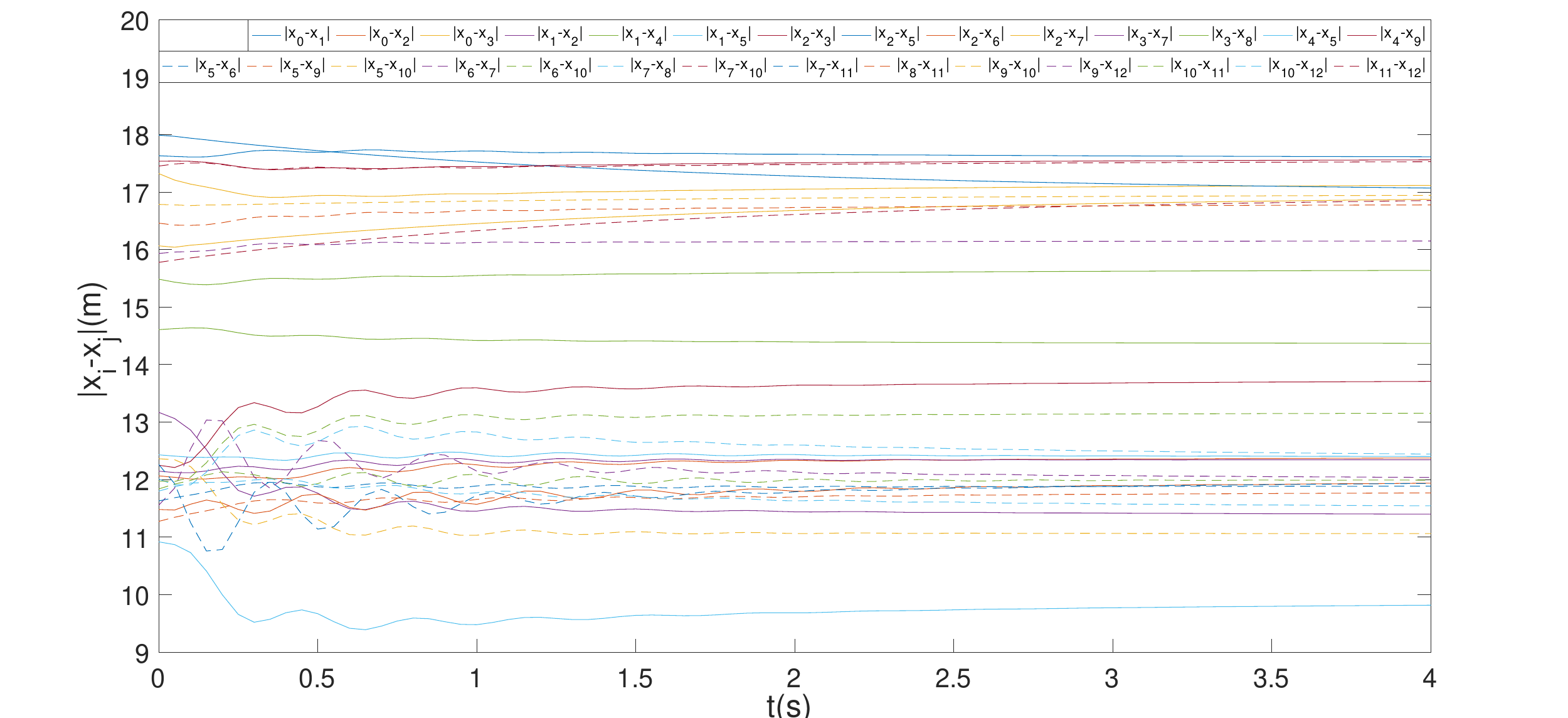}}
\setlength{\abovecaptionskip}{-8pt}
\setlength{\belowcaptionskip}{-8pt}
\caption{Trajectories of distances between the neighboring UAVs.}
\end{figure}

\begin{figure}[!ht]
\vspace{-0.3cm}
\begin{center}
{\includegraphics[width=0.37\textwidth]{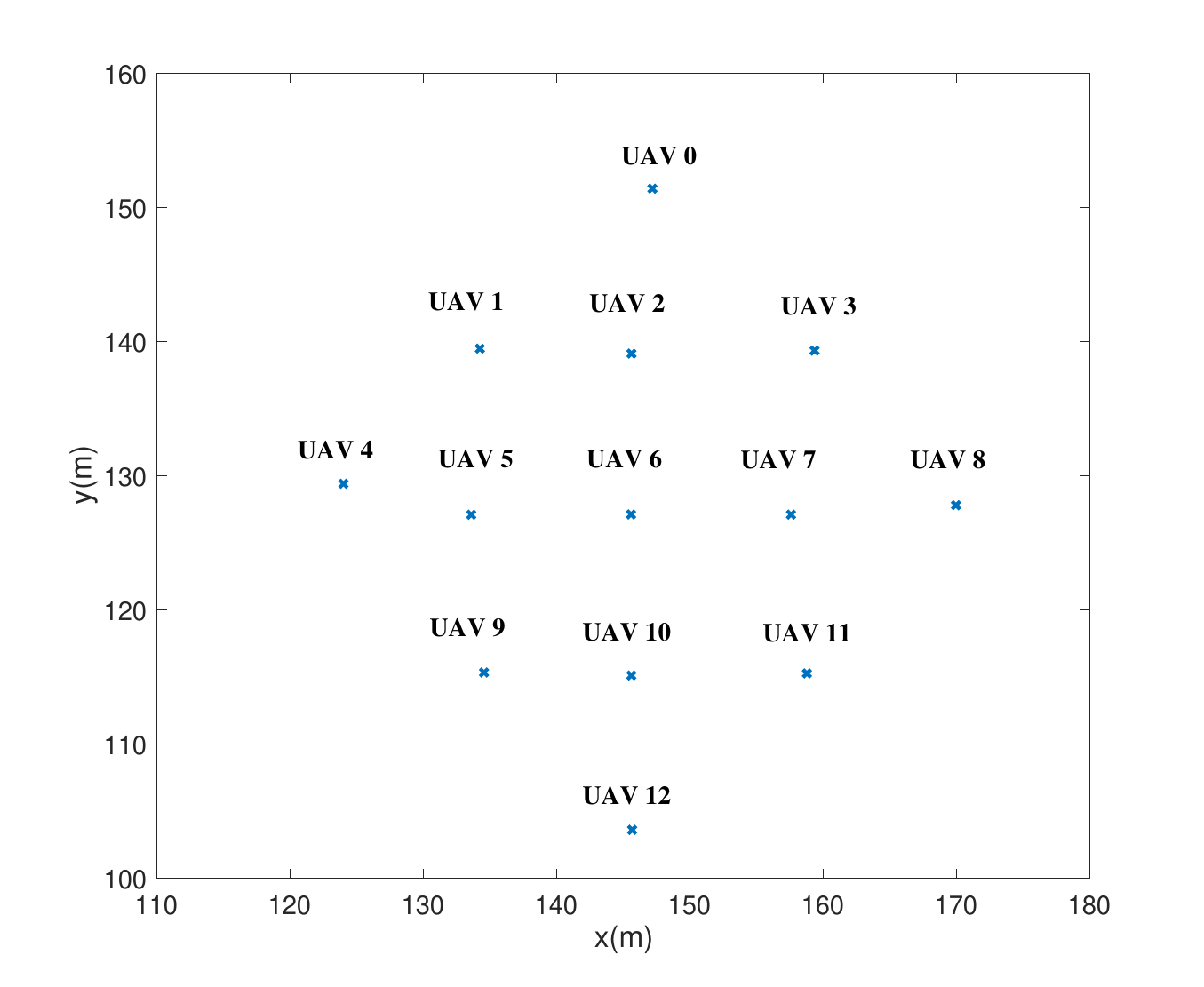}}
\end{center}
\vspace{-0.3cm}
\caption{Flocking patterns at $t=4s$ where $\mathrm{x}$ and $\mathrm{y}$ present positions in $\mathrm{x}$ and $\mathrm{y}$ dimensions, respectively.}
\end{figure}

\section{Conclusion}
This paper, for the first time, considers the flocking control with a malicious agent, and the proposed hierarchical geometric configuration based flocking control method is applied to a swarm with a malicious agent. The new result enriches the conventional flocking control theory.
In the future, by combining the switching system theory and the proposed parameter estimation framework, the malicious agent with changeable parameters will be taken into consideration. Moreover, this new result will be extended to more cases:
one malicious agent acts selectively on a part of its neighbors,
or multiple malicious agents existing in the swarm.
Further studies will also focus on the application of the containment method to multi-agent with nonlinear or other complex dynamics.

\end{document}